\documentclass[10pt]{article}

\usepackage{amssymb,amsmath,bbm}
\usepackage{url}
\usepackage{xcolor}

\usepackage[margin=2cm]{geometry}

\usepackage{natbib}
\usepackage{graphicx}

\PassOptionsToPackage{hyphens}{url}\usepackage{hyperref}

\begin{document}

\title{A spatial agent based model for simulating and optimizing networked eco-industrial systems}
\date{}

\author{J. Raimbault$^1$, J. Broere$^{2,\ast}$, M. Somveille$^3$, J. M. Serna$^4$, E. Strombom$^5$, C. Moore$^6$, B. Zhu$^7$, L. Sugar$^8$\medskip\\\small
$^1$ CASA, UCL, London, United Kingdom\\\small
$^2$ Utrecht University, Centre for Complex Systems Studies, Utrecht, the Netherlands\\\small
$^3$ University of Oxford, Edward Grey Institute, Department of Zoology, Oxford, United Kingdom\\\small
$^4$ Universit{\'e} de Paris, CRPMS, F-75013 Paris, France and PCPP, F-92100 Boulogne-Billancourt, France\\\small
$^5$ University of Minnesota, CBS Ecology, Evolution and Behavior, Minnesota, USA\\\small
$^6$ University of Oxford, Environmental Change Institute, Oxford, United Kingdom\\\small
$^7$ Delft University of Technology, Department of Engineering Systems and Services, Delft, the Netherlands\\\small
$^8$ University of Toronto, Department of Civil Engineering, Toronto, Canada\medskip\\\small
$\ast$ Corresponding author: jorisbroere@gmail.com
}

\maketitle

\begin{abstract}
Industrial symbiosis involves creating integrated cycles of by-products and waste between networks of industrial actors in order to maximize economic value, while at the same time minimizing environmental strain. In such a network, the global environmental strain is no longer equal to the sum of the environmental strain of the individual actors, but it is dependent on how well the network performs as a whole. The development of methods to understand, manage or optimize such networks remains an open issue. In this paper we put forward a simulation model of by-product flow between industrial actors. The goal is to introduce a method for modelling symbiotic exchanges from a macro perspective. The model takes into account the effect of two main mechanisms on a multi-objective optimization of symbiotic processes. First it allows us to study the effect of geographical properties of the economic system, said differently, where actors are divided in space. Second, it allows us to study the effect of clustering complementary actors together as a function of distance, by means of a spatial correlation between the actors’ by-products. Our simulations unveil patterns that are relevant for macro-level policy. First, our results show that the geographical properties are an important factor for the macro performance of symbiotic processes. Second, spatial correlations, which can be interpreted as planned clusters such as Eco-industrial parks, can lead to a very effective macro performance, but only if these are strictly implemented. Finally, we provide a proof of concept by comparing the model to real world data from the European Pollutant Release and Transfer Register database using georeferencing of the companies in the dataset. This work opens up research opportunities in interactive data-driven models and platforms to support real-world implementation of industrial symbiosis.\medskip
\end{abstract}

%%%%%%%%%%%%%%%%%%
\section{Introduction}

The primary mode of production and consumption globally follows the linear pattern of take, make and dispose. Not only does this approach produce an enormous amount of waste, it depletes limited resources while polluting the natural environment and threatening the foundations that support life on Earth~\citep{macarthur2013towards}. Despite scientific consensus on the need for transformation in the production/ consumption sector, few firms are acting with a sense of urgency. The substantial risks associated with maintaining the linear ‘status quo’ for individual firms, whose singular impact may seem insignificant, far outweigh those of pioneering the structural change needed to reduce the impact of production systems globally. However, this status quo can be challenged by shifting to a nonlinear systems perspective, whereby rapid social change is possible and the risks of business-as-usual practices carry catastrophic consequences.

A prominent example of a nonlinear system is what is called an ‘industrial symbiosis’. Industrial symbioses contain mechanisms by which traditionally separate industries work together to reuse and recycle energy, water and by-products, thereby minimizing environmental impact and ideally creating competitive advantage~\citep{chertow2000industrial}. Industrial symbioses are related to the circular economy, which seeks to reduce the uptake of raw materials and reduce total waste production by cycling materials through different uses to maximize the service provided~\citep{macarthur2013towards,geissdoerfer2017circular}. This concept is distinct from traditional recycling where by-products are often reduced to their lowest value raw material form. Circular economies, however, aim to maintain the value-added status of products for as long as possible. This is achieved by sharing the use of products and cycling the highest value form of products and by-pro- ducts as inputs to other firms, which limits the sequential downgrading of by-product material or value.

Integrating by-products and business practices among a multiplicity of previously independent agents creates a complex system. The complexity of such a system is further compounded when considering the inherent social, environmental and economic aspects of production systems. An important factor for realizing functional exchanges is the geographical proximity of other industrial actors~\citep{ghisellini2016review,jensen2011quantifying}. For example, excess heat dissipates more the farther it is transferred. As such, symbiotic activities and optimizing waste flows will be most successful if the industrial actors are close to each other–not only for optimizing physical flows, but also for co- ordinating activities. Eco-industrial parks are considered functional realizations of the industrial symbiotic approach~\citep{gibbs2007reflections}. In these parks, businesses work together to reduce waste and pollution, effectively sharing and exchanging different kinds of resources, infrastructure and by-products. The key advantage is that these actors are located together, thus facilitating these exchanges.

In this study we recognize two important factors for realizing functional and effective symbiotic exchanges. First, to effectively evaluate symbiotic exchanges it is not enough to look at individual exchanges. In order to understand and optimize symbiotic exchanges in terms of sustainability, a system or macro level perspective is required~\citep{chertow2012organizing,martin2015gets}. Here optimal means the theoretically most sustainable symbiotic system, given a set of actors located in a geographical setting. Second, an optimal performing system of symbiotic exchanges may very much depend on geographical features~\citep{desrochers2001cities}, such as where the actors are located in geographical space and which actor is engaging with which actor in realizing symbiotic exchanges. We argue that the first and second point are inevitably connected. Although these two factors are well known in the literature on industrial symbiotic exchanges, the interplay of the two factors have not been studied in detail on a macro- level scale. We argue that there is potentially a lot to gain from studying symbiotic systems as a whole in their geographical context.

Agent-based models have been used when there is the need to model the dynamics of circular economies and industrial symbiosis networks. In an agent-based model (ABM), actors (or agents) interact using prescribed rules, and the emergent behavior of the system is observed~\citep{farmer2009economy}. ABMs have been used to simulate various industrial symbiosis systems, including biogas production~\citep{Yazan2018820}, agro-food systems~\citep{Fernandez-Mena2016467}, agro-industrial complexes~\citep{Huang2015301}, and coal-based industrial systems~\citep{Wang2017636, Wang2014690}, to name a few examples. Simulation methods, such as ABMs, are particularly useful because they can be connected to empirical data and offer flexibility for the processes that can be included in the model~\citep{farmer2009economy,axtell2001agent}. For example, \cite{tsekeris2011spatial} developed a conceptual ABM to simulate the location choice of firms in urban settings and the resulting circular economy flow. \cite{zhu2013exploring} used an ABM to show that strong dependencies between companies is detrimental to the resilience of the system. Other studies have incorporated ABMs into life cycle analyses of products~\citep{wu2017agent,davis2009integration} or into the design process for eco-industrial parks~\citep{Lange2017, Romero2014394, Batten2009197}, while others advocate that the usage of ABMs in the field of industrial ecology is crucial to manage its complex issues~\citep{kraines2006applying,dijkema2009complexity}. Another benefit of using ABMs is their ability to model and anticipate the role of specific policies and technologies on the emergence of industrial symbiosis networks, including the role of information sharing platforms~\citep{Fraccascia2018473}, trust and knowledge diffusion~\citep{ghali2017agent, romero2013framework}, taxes and subsidies~\citep{Fraccascia2017}, institutional capacity building~\citep{Zheng2017}, and different contractual mechanisms~\citep{Albino20164353}. ABM has also been used as a tool to validate existing industrial symbiosis performance indicators~\citep{CoutoMantese20171652,Mantese2016166}.

So ABMs have been used to study specific mechanisms such as life cycle analyses of products or study characteristics such as resilience. Nonetheless, most of the models reviewed above do not explicitly include spatial components, or when it is the case are applied at large geographical scales (local or urban scale). Thus, there are no studies into how well symbiotic exchanges are being utilized given some set of industrial actors in a geographical area and how a certain geographical area might optimally perform. Indeed, according to \cite{velenturf2016promoting}, the question of the role of geographical proximity in industrial symbiosis remains to be investigated. While \cite{chertow2012organizing} propose a theoretical model for the emergence of industrial symbiosis networks at this regional scale, they do not implement it into a simulation model. In this study we aim to fill these gaps by addressing the following research question: How can symbiotic exchanges be optimized in terms of sustainability, given a set of actors located in a geographical area? To address this question, the goal of the current study is to put forward the basis of an agent-based model that can eventually evaluate theoretical optimal symbiotic exchanges given a set of actors in a geographical area and compare it to empirical or alternative scenarios.

The basic model presented in the current study includes two main features. First, the model studies what the effect is of geographical properties on symbiotic relationship by multi-objective optimization, minimizing both cost and waste products. The system size in this case can be varied from regional, to national, to even global. Second, the model studies what the effect is of matching complementary actors in space on the multi-objective optimization, minimizing both cost and waste products. The effect of matching complementary actors is studied by adding spatial correlation between the by-products of actors. This means that the higher the spatial correlation, the more likely that actors close together in space match on their by-products and therefore can engage in a symbiotic exchange. Answers to these questions can help to understand how symbiotic linkages can be optimized given the geo- graphical properties of an area. This macro perspective can potentially be used for policy planning and sustainable urban planning from a macro perspective. Note that policies can rely on top-down approaches (planning) or on bottom-up processes (incentives to change the behavior at the company level)~\citep{velenturf2016promoting}, and both aspects can be investigated with our model.

Our contribution is focused on the effect of geographical proximity on the function of a symbiotic system as a whole, by means of an ABM inspired from ecological concepts, drawing on the transfer of concepts and models between ecology and industrial ecology~\citep{hess2010ecosystem}. Furthermore, this paper addresses the dynamics of how symbiotic exchanges between enterprises are established. In that context, we introduce a model whose linkages must grow organically, thus based on mutual benefit and geographical proximity, and not solely on central control. With fairly simple decision-making processes, a variety of network structures can be formed with implications for the macro-level properties of the system, including system potential to attain circularity goals. By trying to understand the production of macroscopic patterns from the bottom-up, this approach takes into account developments from generative social science~\citep{epstein2006generative} and pattern oriented modeling in ecology~\citep{grimm2005pattern}. It also relates to geosimulation~\citep{benenson2004geosimulation} by simulating spatialized socio-economic processes. The model's agents are enterprises located on a spatial plane, each with an input and an output in terms of needs and waste. All agents have the same goal of minimizing their waste and maximizing their economic profit. Although the contribution is mainly methodological, we present preliminary results and provide a proof of concept by comparing the model to real world data from the European Pollutant Release and Transfer Register database. Our contributions are as follows: (i) to the best of our knowledge, we introduce the first spatial model for growing a symbiotic system at this scale, (ii) we apply state-of-the-art model exploration and calibration techniques with high performance computing to extract knowledge on model behavior, (iii) we therein identify stylized findings from model simulations that may have important implications for policy planning, and (iv) we show that the model can be applied and calibrated on a real world setting. In future work the model will be extended to be more data driven. Therefore, this work introduces a framework that can be extended for studying both practical and theoretical questions.

In this paper, we present the basic rationale of the model, an exploration of its possible applications, and our simulation results along with a test of the model with real world data. The paper is organized as follows. First, we study whether there is a spatial effect on the functioning of the system by comparing a uniform spatial distribution with a theoretical real world distribution and an empirical distribution. Second, we study the effect of geographically matching the actors by their input and output on the functioning of the system. Third, we test the model using real world data from the European Pollutant Release and Transfer Register database. We finally provide an outline of future avenues for research.

%%%%%%%%%%%%%%%%%%
\section{Model description} \label{sec:modelDescription}

\subsection{Rationale}

We developed a spatial ABM (i.e., defined in geographical space) in which agents are industrial companies. Each company is characterized by its geographical location, which is assumed to remain constant through time, and a pair of demand and offer functions. The demand function characterizes the input needed for the company to operate, and the offer function is its non-commercialized output (e.g., water, heat, materials, etc.). In our model, the interaction between two companies, e.g., Company A and Company B, corresponds to Company A buying the waste or by-product (i.e., the non-commercialized output) from Company B to use it as input for its production. The two companies can potentially interact simply based on the geographical distance separating them and the match between input and output (i.e., demand and offer).

Our modeling approach takes an interdisciplinary insight by bridging with the ecological literature, in particular by drawing from the probabilistic niche model (PNM), which can reproduce the structure of complex food webs~\citep{williams2000simple,williams2010probabilistic}. This approach is relevant since the transfer of ecological models to circular economy have already been shown to have a high potential in providing alternative heuristics to optimize recycling systems~\citep{ryen2018ecological}. In the PNM, predation interaction between two species is modeled as the probability of species $i$ eating another species $j$ based on their values along a one-dimensional 'niche' axis. More specifically, species $i$ has a feeding optimum value on the niche axis and the probability of eating species $j$ declines as the niche position of species $j$ gets further from this feeding optimum, which was modeled using a Gaussian centered on the feeding optimum. In our model of industrial symbiosis, we replaced the one-dimensional niche space with a one-dimensional 'by-product' space (which could later be generalized to a multi-dimensional 'by-product' space) along which the input and output functions for each company are defined as Gaussians. The output Gaussian characterizes the uncertainty and variability around a company's by-product, and the input Gaussian characterizes the range of by-product that a company is looking for around a by-product optimum. Our model also extends the analogy with the PNM by being spatial whereby the geographical distance between agents has an effect on their potential interaction, this on potentially large spatial extents which correspond to regional systems in our application. Simulations start with a set of companies that are not linked with each other, and the interaction network then grows. The model focuses on the exchange of by-products between companies and ignores the input of external resources from outside the system (e.g., raw material from the environment), as well as the effective products that companies make for commercialization, which are outputs of the system. It results in a directed network between companies in a closed system, as a macroscopic emerging pattern from bottom-up interactions between companies.

The model includes five distinct processes. Two of them affect the setup of the system: (i) the process governing the geographical positioning of companies, and (ii) a spatial process affecting the position of the input Gaussian of companies along the by-products axis based on the position of the output Gaussian of other nearby companies (in other words, a spatial correlation between the positions of the input and output Gaussians of different companies along the by-products axis). The other three processes affect the growth of the network of symbiotic relationships: (iii) the threshold above which two companies can make a contract to exchange by-products based on how well the respective inputs and outputs of these companies match, (iv) the probability that two companies can interact (i.e., exchange by-products) depends on the geographical distance separating them, and (v) the cost of transporting a by-product from the location of the company selling it to the location of the company buying it.

%%%%%%%%%%%%%%%%%%
\subsection{Model setup}

Companies are indexed by $1\leq i\leq N$, each having a fixed spatial position $\vec{x}_i$, and by-products are described by a finite-dimensional real variable $\vec{y}\in \mathbb{R}^d$ that allows to normalize along each axis and take $\vec{y} \in \left[0,1\right]^d$. A company's demand and offer functions are defined in a simple manner by $\vec{D}_i (\vec{y})= D_i^{(0)}\cdot \vec{d}_i (\vec{y})$ and $\vec{O}_i (\vec{y})= O_i^{(0)}\cdot \vec{o}_i (\vec{y})$ respectively, where $\vec{d}_i$ and $\vec{o}_i$ are multivariate probability densities. For the sake of simplicity, we assume one-dimensional products. In the default setup, these probabilities are fixed as Gaussian distributions with a uniformly distributed average in $\left[0;1\right]$ and a standard deviation given by a parameter $\sigma$.

%%%%%%%%%%%%%%%%%
\begin{figure*}
\includegraphics[width=\textwidth]{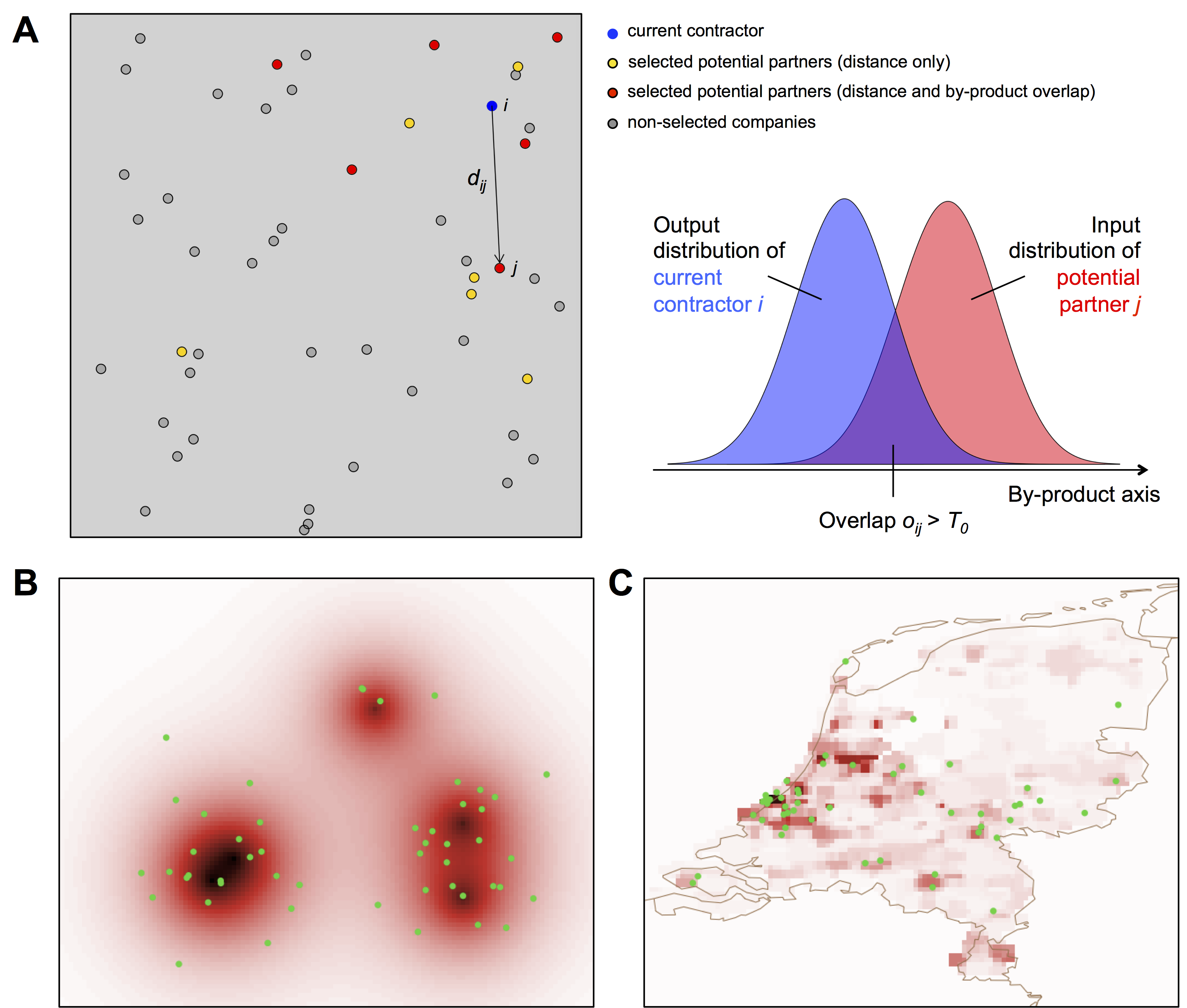}
\caption{\textbf{Model setup and illustration.} (A) Example of a random distribution of companies, represented by points on a uniform space. First, one current contractor \textit{i} (blue point) is selected among the companies. Then, a first set of potential partners for this current contractor is sampled based on distance (i.e. geographical interaction potential; yellow and red points). Among this set, potential partners whose input density distribution overlaps with the current contractor's output density distribution more than the threshold $T_o$ (see illustration on the right-hand side) are retained (red points). The bottom plots show examples for two possible geographical setups: a synthetic city system (B), and real density data for the Netherlands (C), in which green points indicate the location of companies generated stochastically based on population density.}
\label{fig:Variables}
\end{figure*}
%%%%%%%%%%%%%%%%%

\paragraph{Geographical distribution of companies.}
We setup the initial geographical position of companies in four different ways: 

\begin{itemize}
\item Using a spatial uniform distribution for the geographical coordinates (illustrated in Figure~\ref{fig:Variables}A).
\item Simulating a synthetic population density field $d(\vec{x})$, and assuming that the number of companies $Y$ follows a scaling law as a function of the population of a city $N$ such that $Y \sim N^{\beta}$. Industries naturally do not locate in the middle of large cities, but at the aggregated scale considered, high population densities and industries aggregate in built-up areas and the number of industries in cities do follow a scaling law of population~\citep{pumain2006evolutionary}.
More specifically, we took the probability for a firm to locate in a patch as a function of its population 
\begin{equation}
\mathbb{P}\left(\vec{x}_i=\vec{x}|i\right) \propto \left(\frac{N(\vec{x}}{\sum N}\right)^{\beta}
\end{equation}
Companies are thus located sequentially at random, given these probabilities, and the population distribution is synthetically generated~\citep{raimbault2019urban}, as a kernel mixture
\begin{equation}
P(\vec{x}) = \sum_{1\leq j\leq p} K_j (\vec{x})
\end{equation}
with $p$ number of cities (or ``centers''), with kernels
\begin{equation}
K_j(\vec{x}) = \cdot \exp{-\frac{||\vec{x}-\vec{x}_j||}{r_0}}
\end{equation}
where $x_j$ is random with an uniform distribution and $r_0$ is computed such that the city system respects the Zipf rank-size law with exponent $\gamma$ (similar values at origin assume a constant maximal center density across cities), i.e. such that $P_j = \iint K_j \propto \frac{1}{j^{\gamma}}$. This way of generating a geographical distribution of companies is illustrated in Figure~\ref{fig:Variables}B.
\item Using a real raster describing the geographical distribution of population density across the Netherlands with 1km resolution (data from ~\cite{gridded2005density}; illustrated in Figure~\ref{fig:Variables}C). This shows the typical scale considered for the system of cities. Companies were then positioned as above.
\item Using a real deterministic system of companies' geographical positions. The application of the model to a real case below shows such a setup for Netherlands in Fig.~\ref{fig:eprtr}. An illustration at a local scale for the Kalundborg Eco-industrial Park in Denmark is also given in Appendix in Fig.~\ref{fig:RealExam}.
\end{itemize}

The simple setup through population distribution for the positioning of companies, except for the last case, allows to be flexible and for example compare the effects due to model dynamics to the effects due to geography.

\paragraph{Spatial correlation for demand-offer functions.} 

A complementary setup mechanism is included in the model to directly test the implications of policies such as the implementation of industrial parks \citep{su2013review}. We abstract such processes by controlling on correlations between the industrial production structure of companies. More precisely, the averages of Gaussian input and output distributions are not uniformly distributed, but are a function of the position in space such that a certain level of correlation is controlled by a parameter.

This level of correlation is expected to be a function of the distance between companies, corresponding to policies locating companies which can fit closer together. However, generating random variables located in space and which correlation matrix is their distance matrix is generally not possible, and relates to a distance matrix completion problem: to be a correlation matrix, a given matrix must be symmetric positive definite (the Choleski decomposition giving then the transformation from independent drawings to the correlated variables)~\citep{bakonyi1995euclidian}. We adopt thus an heuristic strategy based on mean fields and that furthermore have a direct thematic interpretation.

We assume to be in the synthetic city system setup or in the real world setup, for which both are defined urban centers. We associate randomly to each center values that will determine the distribution averages within their geographical span. These values are taken as evenly spaced within $\left[0;1\right]$ to maximize the discrepancy between them. Given a new company, its distribution average will be given by a Gaussian distribution following $\mathbb{E}\left[o_i\right] \sim \mathcal{N}(a_c,\alpha \cdot d) $ where $a_c$ is the value of the distribution average of the closest center, $d$ is the distance to this center normalized by the maximal distance (so that $d \leq 1$) and $\alpha$ is the parameter controlling the local variability of averages.

Taking $\alpha = 0$ induces uniform averages within Thiessen polygons of centers, and thus a high spatial correlation between averages, whereas large values of $\alpha$ will make them uncorrelated. It can be interpreted as a `level of clustering', that can be acted upon through policies, by fostering local correlations between the type of production of companies (we recall that in our abstract model, the distribution of $o_i$ and $d_i$ describe the type of input needed by a company and the type of output it produces).

%%%%%%%%%%%%%%%%%%
\subsection{Growing the network of symbiotic relationships}

Once the geographical position of companies is set up, the urban environment (which includes the transportation cost landscape) is assumed to remain constant, and the simulation of the network growth (i.e. the progressive establishment of complementary links between companies that correspond to flows of by-products) can start. Two factors are used to establish links between companies (i.e. exchange of by-products): (1) the geographical distance separating them, and (2) the match between demand and offer, and these links are added sequentially. At each time step, the following set of rules is applied in sequence to establish a link (and therefore grow the network of symbiotic relationships): 

\begin{itemize}

\item A company, the `current contractor', is drawn at random uniformly among the companies with minimal number of links (we recall that a link between two companies corresponds to a contract settled in a previous step, and consider therefore that companies with minimal exchange activity will try to make new contracts).

\item A first sample for potential partners for this company is drawn based on a geographical interaction potential ($V_{ij}$; i.e. probability that the current contractor $i$ and another company $j$ interact based on their geographical locations), which is defined as
\begin{equation}
V_{ij}=\exp{\left(-d_{ij}/d_0\right)}
\end{equation}
where $d_{ij}$ is the geographical distance between the two companies and $d_0$ is the characteristic geographical range within which potential partners are typically drawn.

\item For each company previously sampled, the overlap between the current contractor's offer (i.e. what it wastes after production) and the company's demand (i.e. what they could use for production), both represented by Gaussian density distributions along the by-product axis (see Figure~\ref{fig:Variables}A) is computed as
\begin{equation}
o = \int min(O,D) dx
\end{equation}
with a higher overlap indicating a higher probability that the two companies exchange by-products. Companies whose overlap is above $T_o$ are taken as potential partners.

\item For each potential partner $j$, a utility associated with the potential exchange of by-products with the current contractor $i$ is computed as 
\begin{equation}
u_{ij} = o_{ij} - c \cdot \frac{d_{ij}}{d_{max}}
\end{equation}
where $o_{ij}$ is the overlap between the two companies in by-product space, $c$ is the transportation cost that is assumed to be shared between the two companies, $d_{ij}$ is the geographical distance between the two companies, and $d_{max}$ is the maximum distance between any two companies in the system. Then, given the set of utilities $(u_{1j},u_{j1})_j \simeq (u_j)_j$, the potential partner with best utility is chosen.

\item The current contractor's Gaussian offer distribution and the partner's Gaussian demand distribution are updated by truncation of the overlap area, as it is not available anymore for exchanges with other companies.

\end{itemize}

The growth of the network of symbiotic relationships ran until the cumulated variation of output waste becomes negligible and after a minimal number of iterations. More precisely, if $w_{j,t}$ is the output waste for company $j$ at time $t$, the stopping criteria is $\frac{1}{N} \left|\sum_{j} w_{j,t} - w_{j,t-1}\right| < \varepsilon$ where we took $\varepsilon = 10^{-3}$. In practice, all simulations satisfied this criteria at $t_f = 500$.

\begin{figure*}
    \includegraphics[width=\textwidth]{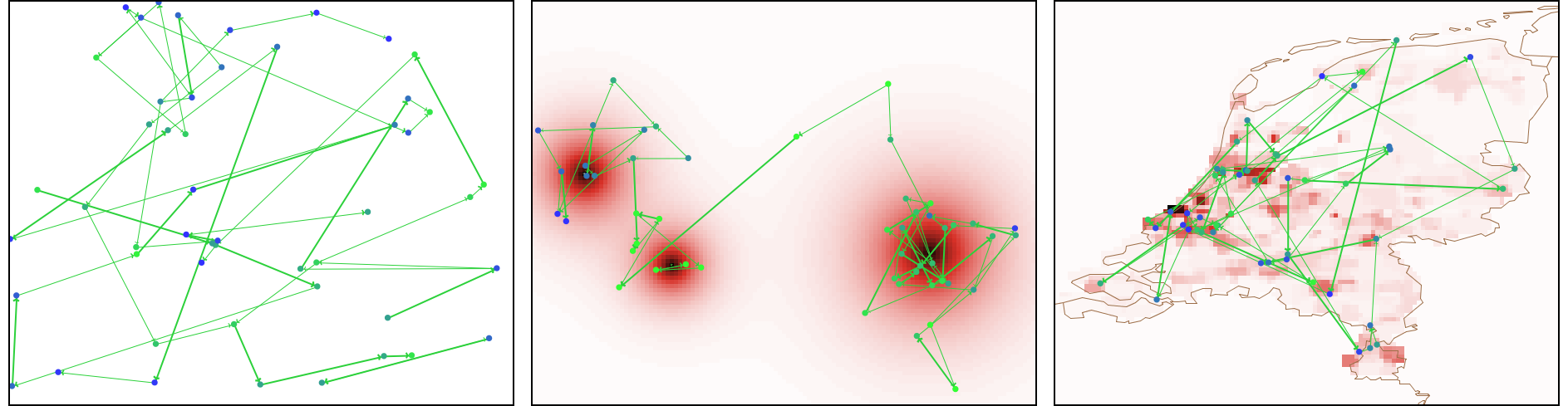}
    \caption{\textbf{Examples of networks generated.} We show final networks for the three different geographical setups (uniform, synthetic city system, actual population of The Netherlands). Parameters are set to $d_0 = 40$, $T_0 = 0.06$, $c = 1.3$, $\sigma = 0.04$, $\alpha = 2.4$, and the process to correlate averages is active in the synthetic city system only, where centers are defined. We already visually observe significant differences in the form of the final network depending on the geography.}
    \label{fig:examples_final_network}
\end{figure*}

To visualize the network between companies and give an intuition of this stopping criteria, we illustrate in Fig.~\ref{fig:examples_final_network} final configurations obtained for the same parameter values but with the different geographical setups.

%%%%%%%%%%%%%%
%%%%%%%%%%%%%

\subsection{Model outputs}

Given an economic configuration generated by the model, we need indicators to quantify its performance at an aggregated level. The different objectives, such as sustainability that has to be maximized and cost that has to be minimized, are generally contradictory for such complex economic systems~\citep{fazlollahi2012methods}.

We use simply the following indicators to quantify the performance of a final configuration:
\begin{itemize}
	\item Total waste $W$, which is the sum of remaining areas for all output distributions at final time. Thus, total waste corresponds to the quantity of waste which was not exchanged by companies.
    \item Relative cost $C$, given by the total length of the network (sum of lengths of all links) weighted by flows within links, normalized by world scale (diagonal of the world). Note that we do not include transportation cost in the computation, so that costs for different values of $c$ can be compared. In other words, this relative cost is the effective total cost of moving materials normalized by transportation cost.
\end{itemize}

%%%%%%%%%%%%%%%%%%
\subsection{Model summary and implementation}

The central feature of agent-based models is that they evolve macroscopic patterns \emph{from the bottom-up}, namely with assumptions done at the microscopic level on the characteristics of agents and their behavior, with few top-down macroscopic constraints (in our case these are for example locations and transportation cost, while company product distributions, partner choices, etc. are done at the microscopic level). The simulation model starts from an initial configuration (here companies with no symbiotic link) and simulates step-by-step the evolution of the system following the simplified rules for company behavior. With time flowing, companies will successively choose partners and create symbiotic links. The final network of relations depends on the realized sequence of events, and therefore on random fluctuations which are also simulated, but as we show with model exploration and statistical analysis, macroscopic indicators (global network structure indicators, including total cost, total final waste), are statistically robust as a function of microscopic model parameters. The cost of a relation for each link is the travelled length weighted by the waste quantity, which allows computing the total cost, and the total waste, in a similar way. This emerging behavior can not be predicted other than through simulation. This also explains the stylized experiments we will conduct on the model, like studying the influence of transportation cost which can directly be interpreted in the real world, or trying to optimize total cost and waste, or studying the impact of spatial correlations, since the simulation model can be used as a ``virtual laboratory'' to test hypotheses and scenarios~\citep{epstein2006generative}.

The model was implemented in \texttt{NetLogo}, which is suited for the interactive exploration of such models in which the spatial structure is crucial. We also developed a \texttt{R} version, especially with the objective of an integration into a \texttt{Shiny} web application for a real world use as described before. Model exploration was done using the software OpenMole, which provides simultaneously (i) a seamless embedding of any model in most of existing programming languages into a platform for model experiments; (ii) a transparent access to High Performance Computing environments such as clusters or computation grids; (iii) specific methods for model exploration (e.g. design of experiments, search for diversity, dimensionality reduction) and model optimization and calibration (genetic algorithms) methods~\citep{reuillon2013openmole}. In the study of such a computational model, an intensive exploration is crucial to extract relevant knowledge from it~\citep{banos2013pour}.

Model code and results are available on the open repository of the project at\\\url{https://github.com/SFICSSS16-CircularEconomy/CircularEconomy}. Large simulation results files are available on the dataverse at \url{https://doi.org/10.7910/DVN/7XCWTN}.

%%%%%%%%%%%%%%%%%%
\section{Results}

Model experiments were run with varying the parameters $d_0, \sigma, T_0, c$, and the type of setup between uniform and synthetic city system for a baseline experiment, as well as varying $\alpha$ on a targeted experiment on a synthetic city system. The meta-parameters (parameters for model geographical initialization which remain fixed in experiments) were fixed at $N=50$ (one sector of reasonable size), $\gamma = 1.3$ (yielding a hierarchical city system~\citep{10.1371/journal.pone.0183919}), $\beta = 1.5$ (supra-linear scaling law between population and industry, corresponding to a high added-value sector~\citep{bettencourt2007growth}) and $p=5$ centers within a square world of size 100km (typical of a regional city system as shown for example in the case of France by~\cite{berroir2017systemes}, and being the magnitude of the application to Netherlands). Boundaries of varying parameters and details of numerical experiments are given in~\ref{app:experiments}. The choice of fixing these setup parameters aims at studying the dynamics of the model without capturing a variability due to geography (which could be a possible development as done by \cite{raimbault2018space}).

We recall here, as detailed in Table~\ref{tab:params} with the associated processes, that the varying parameters are the transportation cost $c$ and the gravity decay $d_0$ (both linked to spatial interactions), the threshold $T_0$ and the distribution width $\sigma$ (both linked to industrial structure) and the correlation level $\alpha$ (linked to industrial clusters). Studying the role of each thus informs on the corresponding process. This type of model validation through statistical consistency, sensitivity analysis, and calibration (here with a bi-objective minimization of stylized objectives as detailed below), is typical for such simple simulation models~\citep{pumain2017urban}.

%%%%%%%%%%%%%%%%
\begin{table*}[h!]
\caption{\textbf{Summary of model parameters.} We give first meta-parameters that are fixed during our experiments, and parameters that varied during experiments. We give the corresponding process, the range and the default value for each. All parameters are adimensional, except gravity decay which is in kilometers.}\label{tab:params}
	\begin{tabular}{|l|l|l|l|l|l|}
	\hline
	Parameter & Notation & Process & Range & Value & Observation\\ \hline
	Number of firms & $N$ & Economic system & $[2 ; 10^6]$ & $N = 50$ & Upper bound depends of scale\\ 
	Hierarchy of city system & $\gamma$ & City system & $[0.5 ; 2.0]$ & $\gamma = 1.3$ & - \\ 
	Density-to-firms exponent & $\alpha$ & Economic system & $[0.1 ; 4.0]$ & $\alpha = 1.5$ & -\\
	Number of centers & $p$ & City system & $[1 ; 10]$ & $p = 5$ & - \\\hline
	Gravity decay & $d_0$ & Spatial interactions & $[1;200]$ & $d_0 = 50 km$ & Depends on scale \\
	Distribution width & $\sigma$ & Industrial structure & $[0.01 ; 0.1]$ & $\sigma = 0.05$ & Candidate for policies \\
	Overlap threshold & $T_0$ & Industrial structure & $[0.01 ; 0.1]$ & $T_0 = 0.1$ & Candidate for policies \\
	Transportation cost & $c$ & Urban system & $[0.1 ; 4.0]$ & $c = 0.5$ & Exogenous \\
	Correlation level & $\alpha$ & Industrial clusters & $[0 ; 20.0]$ & $\alpha = 5$ & Candidate for policies \\\hline
	\end{tabular}
\end{table*}
%%%%%%%%%%%%%

%%%%%%%%%%%%%%%%%%
\subsection{Statistical consistency of the model}

First of all, we verify the internal consistence of the model by looking at statistical distribution of indicators (illustrated in~\ref{app:distribs} for some points of the parameter space). Most of the distributions are unimodal but not necessarily normal. However, we can roughly estimate the number of runs needed to reach a certain confidence interval on the mean and be able to differentiate indicators across different parameter values. For example, assuming a normal distribution of the indicator, to obtain a $\alpha$ level confidence interval of width $\frac{\sigma}{2}$ around the mean, the equation $\sigma = \frac{4\sigma \cdot z_{1-\alpha}}{\sqrt{n}}$ must be verified. This leads to $n\simeq 64$ for a 95\% confidence level. For a grid of 11,700 parameter points (see~\ref{app:experiments}) and $n=100$ repetitions, we compute the distance between averages relative to standard deviations, i.e. $r_{ij} = 2 \cdot \frac{\hat{\mu}_i - \hat{\mu}_j}{\hat{\sigma}_i + \hat{\sigma}_j}$ if $\hat{\mu}_k,\hat{\sigma}_k$ are average and standard deviation of the studied indicator for parameter point $k$, estimated on repetitions. This rate $r_{ij}$, computed for all $i \neq j$, summarize if the difference in averages is significantly different from standard deviations, and thus from confidence intervals with the setting we just described. For the waste indicator, the rate $r_{ij}$ has a median of 2.4 and is above 1 at the 22\% quantile, whereas for the cost, the median is at 2.7 and it reaches 1 at 21\%. This number of runs is thus already satisfying for differentiating indicators across parameter variations, and we run experiments with $n=100$ in the following.

%%%%%%%%%%%%%%%%%%
\subsection{Model exploration}
\label{subsec:exploration}

In order to study the baseline behavior of the model, we ran grid search experiments with a uniform initial distribution and a synthetic city system. In~\ref{app:sensitivity} we show figures giving an extensive view of the behavior of indicators. From this exploration we learn the following stylized facts:

\begin{itemize}
	\item Qualitative behavior of indicators that can be intuitively expected: (i) a decrease of waste when $\sigma$ increases (companies with more diverse inputs and outputs have more opportunities to exchange); (ii) an increase of waste when transportation cost increases (potential exchanges which are far in terms of distance become too costly to be realized); (iii) an increase of relative cost when $c$ decreases (lower transportation costs yield more exchanges, but also less efficient exchanges as companies will less seek to optimize their exchange network).
    \item Some behaviors which could not be predicted intuitively from model processes, i.e. corresponding to emergent behavior obtained through simulation: (i) an increase of waste with $T_0$ for low transportation costs (which should correspond to a congestion effect); (ii) a slight u-shape behavior of cost as a function of $\sigma$ for high $d_0$ and low $c$ (when interactions are free in space, intermediate industrial structures are the worse for cost); (iii) an effect of $d_0$ on waste which leads to no difference between the setup types for high values of $d_0$ (gravity decay mitigates the role of space).
    \item The behavior on synthetic city systems and real population data shows different qualitative patterns, confirming the necessity to embed the model in realistic or real data. In other words, \emph{space matters} for the establishment of a circular economy network.
\end{itemize}

%%%%%%%%%%%%%%%%%%
\subsection{Patterns of policy optimization to grow the circular economy}
\label{subsec:optimization}

%%%%%%%%%%%%%%%%%%
\begin{figure*}
	\includegraphics[width=\textwidth]{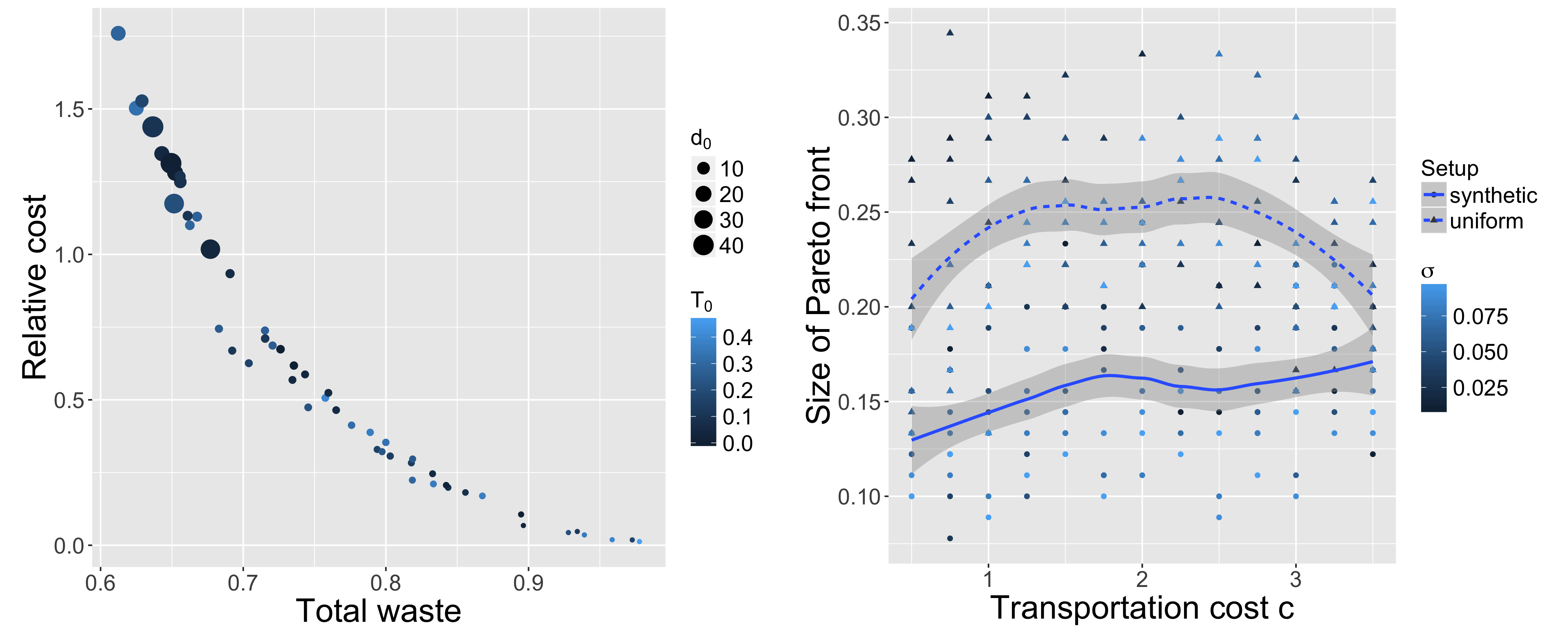}
	\caption{\textbf{Multi-objective optimization of cost and waste.} (\textit{Left plot}) An example of Pareto front, obtained with a NSGA2 algorithm at fixed $\sigma = 0.01$ and $c=3$. (\textit{Right plot}) Relative size of approximate Pareto fronts, defined as the rate between number of points lying on the Pareto front and the number of points in the cloud, as a smoothed function of transportation cost, for both setup types (linetype). Color level gives the value of $\sigma$.}
	\label{fig:pareto-regime}
\end{figure*}
%%%%%%%%%%%%%%%%%%

An abstract application of the model paves the way for the exploration of potential policy optimization. We follow the rationale that the policy makers can influence on some parameters only, under the assumption that: (i) transportation parameters are fixed by exogenous conditions, that include among other factors transportation infrastructure and energy price~\citep{raimbault2017cost} (these aspects are concerned by policies at a different level, both for scope and coverage); and (ii) distribution width is fixed, corresponding to the fixed industrial structure (roughly stylized in our model), which temporal scale of change is significantly larger in magnitude than the one of the model.

In that context, the policy maker can influence the interaction range (gravity decay $d_0$) by giving incentives for collaboration between companies or a better circulation of information for example, and the collaboration threshold $T_0$, also with incentives or technological help. These parameters correspond to relatively easy-to-implement policies in the short term. Therefore, we study optimization patterns on the parameter plan $(d_0,T_0)$, at fixed $(\sigma,c)$, for both objectives simultaneously. We run a model calibration with a genetic algorithm (GA) to demonstrate the existence of a Pareto front of compromise solutions (see~\ref{app:ga} for details). Such a front is shown in Fig.~\ref{fig:pareto-regime} for $\sigma = 0.01$ and $c = 3$. The GA yields a continuous Pareto front without any degeneracy (part for which it would be equivalent to optimize one objective). We furthermore confirm the expected role of $d_0$, for which low values give solutions with low cost and high waste, and high values solutions with low waste but a high cost. The existence of this front has important implication for policy making, confirming that a compromise has to be made between sustainability and the economical cost needed to reach it.

We study then the influence of $(\sigma,c)$ on patterns of optimization. As running the optimization GA has a high computational cost, for this we use the grid search baseline experiment, with the assumption that some information is contained within corresponding approximate Pareto fronts. We show in Fig.~\ref{fig:pareto-regime} the relative size of fronts (defined as the number of non-dominated points relative to the size of the point cloud) when $\sigma$ and $c$ vary. The uniform setup exhibits a u-shape as a function of transportation cost $c$, whereas this relative size is increasing with $c$ for synthetic city systems. As the uniform setup can be understood as a highly zoomed geographical scale, this implies that the highest number of alternatives will occur for intermediate values of transportation cost for local policies. In the case of a city system (regional policy), increasing the exogenous constraint through energy price counter-intuitively increases the number of alternatives for optimization. This means that stronger constraints in fact enlarge the set of Pareto-equivalent choices the decision-maker has to choose within. In terms of quantitative flexibility of objectives (variation of the shape of fronts, with a complex interplay between $\sigma$ and $c$, is shown in~\ref{app:fronts}), we find that the relative spread of fronts, defined as $\frac{\max W - \min W}{\max C - \min C}$, capturing the relative flexibility on each objective, has significant variations from 0.025 to 0.2 when $c$ increases, confirming that changing $c$ yields higher variations on final costs, in line with a higher number of optimization alternatives.

%%%%%%%%%%%%%%%%%%
\subsection{Spatial correlation between input and output distributions}

%%%%%%%%%%%%%%%%%%
\begin{figure*}
	\includegraphics[width=\textwidth]{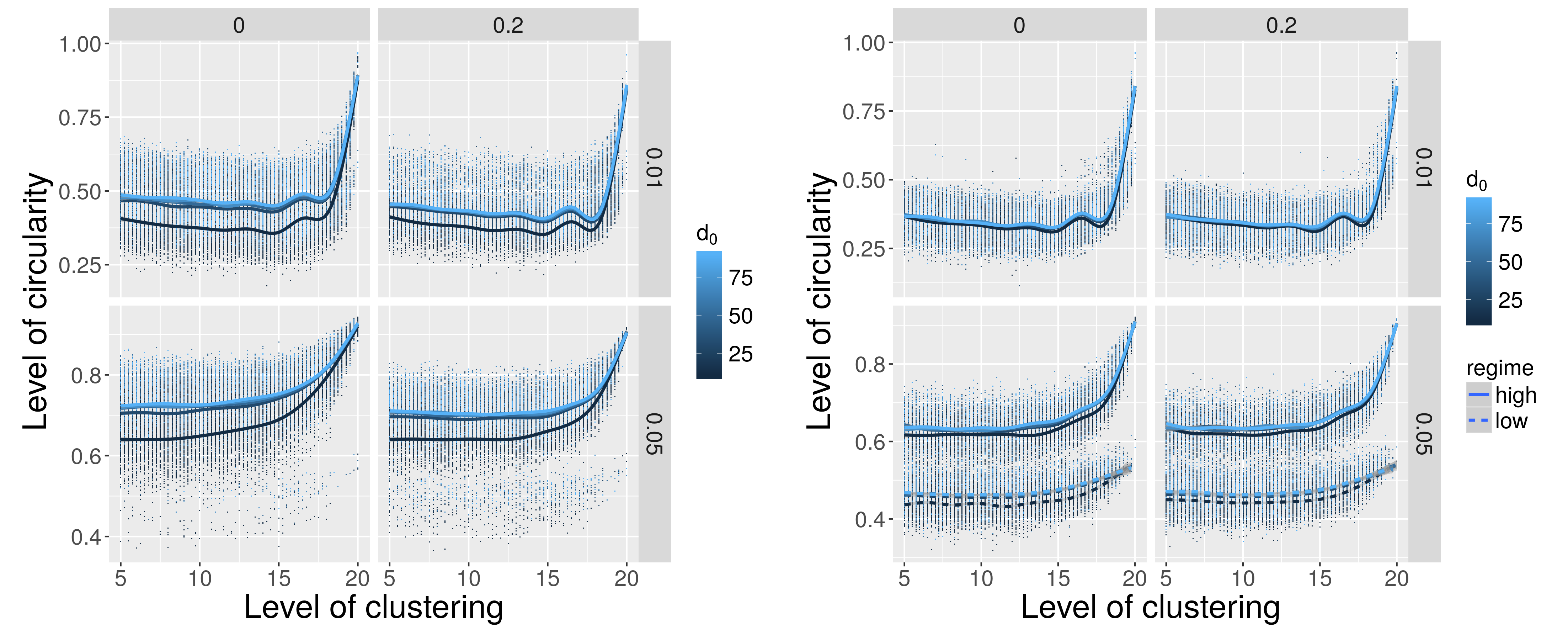}
	\caption{\textbf{Influence of clustering level on the level of circularity.} We plot the level of circularity as a function of the level of clustering, for a low transportation cost $c=0.5$ (\textit{Left plot}) and a high transportation cost (\textit{Right plot}), for different values of the overlap threshold $T_0$ (columns), of the distribution width $\sigma$ (row), and of the distance decay $d_0$ (color). When two regimes clearly emerge (for high $T_0$ and high $\sigma$), they are separated by linetype to fit the summary statistics.}
\label{fig:clusteringlevel}
\end{figure*}
%%%%%%%%%%%%%%%%%%

In Fig.~\ref{fig:clusteringlevel} the influence of clustering demand and offer on the level of circularity is plotted. There is a clear increasing relationship between the clustering on the amount of circularity, indicating that matching actors together in close proximity yields exponentially better results.

For low values of $\sigma$ (here $\sigma = 0.01$), i.e. constrained industrial structure, the gain is significant only for the highest levels of clustering, and even shows a slight decrease and an oscillation before increasing. Although the decrease and the oscillations are statistically significant in terms of confidence intervals, they are negligible regarding the standard deviation of distributions, and so we do not consider them as meaningful. For a high $\sigma$ (here $\sigma = 0.05$), the curves are monotonous but still stagnate for low levels of clustering.

Logically, high transportation costs make the model non-sensitive to the gravity decay $d_0$, and furthermore witness the emergence of a specific mode with a very low performance (dashed lines, fitted separately, these parameter settings correspond to the few bimodal distributions for which the average does not summarizes model modes): in this case, the geographical setting and parameter values produce a very poor network of connections between companies. The existence of this regime is in itself interesting, as it confirms the importance of the spatial configuration.

The most important result for our purpose is that low levels of clustering hardly have any effect on the circularity of the system, as below a high threshold of 15, the circularity remains roughly stable. Above this step, all configurations witness an exponential increase and thus a strong effect of the clustering. These results seem to suggest that matching companies to be located within the same center/city/industrial park can only have a significant effect on circularity if this matching is moderate to strong.

This is an important finding for urban planning and the developments of Eco-industrial parks, as it suggests that a regional view of the industrial system is necessary to optimize the circularity at this level, since local policies will less likely result in such correlation levels for the whole range of industrial products.

%%%%%%%%%%%%%%%%%%
\subsection{Real world deterministic geographical positions}

%%%%%%%%%%%%%%
\begin{figure*}
\centering
\includegraphics[width=\textwidth]{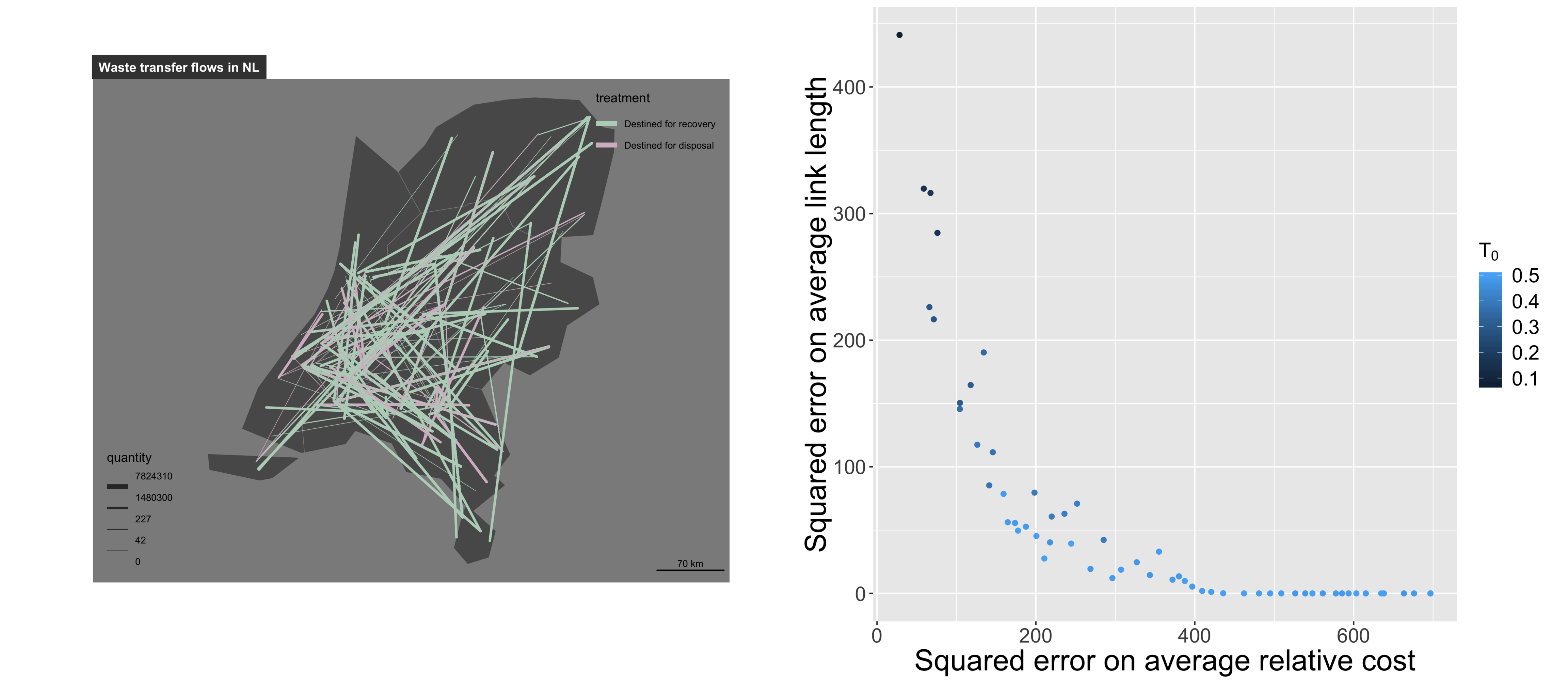}
\caption{Model calibration on the EPRTR database. \textit{(Left)} Map of real links taken into account for calibration. Link width gives the weight which is proportional to real transfer quantity, and link color gives the type of treatment; \textit{(Right)} Conditionally to having an error on network size of less than 10\% of the target value ($N=284$), we show the Pareto front for the remaining objectives of average network length and average network cost. Point color gives the value of calibrated $T_0$.}
\label{fig:eprtr}
\end{figure*}
%%%%%%%%%%%%%%%

We illustrate the potentiality of real-world application of our model at a smaller scale with a large database, more precisely the EPRTR database. Following the regulation 166/2006 of the European Commission, a large proportion of industrial facilities have to declare their pollutant release, but also the transfer of waste to other facilities. The open access database including these declarations is available at \url{https://prtr.eea.europa.eu/}. We use the latest version available (v16) covering the years 2004-2017, which contains information for 62,872 facilities, 9,633 waste handlers and 622,860 waste transfers. After georeferencing addresses in the database, filtering on links with origin and destination available, and considering links inside countries only (for a national scale application of the model), we find that the country with the largest number of links is Netherlands with 284 links (among 1074 intra-country links - most transfers being international a large part of the database is not used at this scale but this number is still enough to apply our model). These links are mapped in Fig.~\ref{fig:eprtr}.

The model is parametrized with real company positions, but no initial links and the same free parameters as before (namely gravity decay $d_0$, distribution width $\sigma$, overlap threshold $T_0$, transportation cost $c$, and no distribution correlation, i.e. $\alpha=0$). We generate the network and compare it to the real links with the following indicators capturing network structure: (i) squared error on number of links $\varepsilon_N$, which can be seen as a preliminary requirement (constrained optimization); (ii) squared error on average link length $\varepsilon_L$ (the average being taken on all links); and (iii) squared error on average relative costs $\varepsilon_C$.

We calibrate the model using the same optimization heuristic as above, i.e. a genetic algorithm run on a computation grid with the software OpenMOLE, until convergence of the result population. We show in Fig.~\ref{fig:eprtr} calibration results (right panel). More precisely, we select the final population points which have a reasonable value for the network size objective, namely with an error less than 10\% of the target value (284 links for the real network). This yields 59 points out of 200. They form a Pareto front regarding the two remaining objectives with relatively low values, witnessing of a good bi-objective constrained calibration (for example, a value of 100 for the error on link length corresponds to a 6\% relative error). We furthermore observe a continuous variation of $T_0$ along this front, high values give the best solutions regarding link length while low values give the best solutions regarding link relative cost: in the case companies have a very low propensity to exchange (high $T_0$) network topology is accurately reproduced, while a higher propensity to exchange yields a better fit on weighted network structure. Considering the compromise points near the origin, such that $\varepsilon < 100$ and $\varepsilon_C < 200$, we can interpret the average estimated parameters: gravity decay $\hat{d_0} = 16.3$ implying mostly local interactions between companies, overlap threshold $\hat{T_0} = 0.46$ which means a high propensity to exchange, transportation cost $\hat{c} = 0.26$ which is relatively low, and distribution width $\hat{\sigma} = 0.01$ corresponding to specialized companies.

We show thus that the model can accurately approach the real network, and the solutions could be applied for the testing of policies, for example running the model with calibrated parameters but changes in spatial distribution of companies. This example is in line with the previous experiments, in which indirect knowledge on processes is gained (e.g., see subsection~\ref{subsec:optimization}). Although it still remains simplified, it shows a proof-of-concept on how the model could be applied to real-world cases, exploiting the insights from real-world data as well as using model simulations.

%%%%%%%%%%%%%%%%
\section{Discussion}

In this paper, we introduced the basis of an ABM for modeling geographical features of industrial symbiotic processes. The main contribution is that the model provides a framework for studying macro-level properties of symbiotic systems given the properties of a geographical area. The model can be extended to accommodate both theoretical and more practical research questions. As a theoretical tool, the model can be used as a formal inductive method to produce hypotheses about complex symbiotic interactions. As a practical tool, the model can help to optimize the effectiveness of symbiotic exchanges given a geographical area. Furthermore, we provide a proof of concept by comparing the model to data from the European Pollutant Release and Transfer Register database, and present some first results. This study is an important step in understanding the effect of geographical features on symbiotic exchanges from a systems perspective, filling a gap in the research on circular economies and industrial symbioses.   

The first general result of this study is that the geographical distribution matters. Although this is known on a local scale, the finding that the behavior is very different between a uniform setup and a synthetic city system is not trivial. Despite the distance between actors being smaller within the city cluster, the distance is bigger for actors between clusters, keeping the average distance the same between setups. The emergent behavior is therefore an interesting phenomena. A consequence of this finding is that optimization of symbiotic exchanges requires custom planning at the local level, but also taking into account the regional or even national level. Secondly, an important finding is that clustering companies that have correlations in their input and output has an exponential effect on the circularity of the system. As a consequence, there is potentially a lot to gain by matching agents together with a specific geographical area. This effect is however attained only above a high threshold of clustering, which means that corresponding policies must be strictly implemented. Spatial correlations also make the local scale less dependent on the regional or national scale for the optimization of the symbiotic system. Therefore the results can be important for policy advice on urban planning and the developments of Eco-industrial areas.

Although this is a simplified model, we obtain less than 10 percent error margin on two network structure indicators when we run the model using georeferencing of companies in the Netherlands and compare the results to data from the European Pollutant Release and Transfer Register. These results are encouraging for the validity of the model in real world applications, and possible future policy applications in a more data-driven way.

Our main contribution consists in developing a methodology for studying industrial symbiotic mechanisms in different geographical contexts. The model will be developed further to accommodate both theoretical and practical research questions. An example of a theoretical question could be to study the effect of different dynamics of industrial symbioses on the performance, such as self-organization, third party mechanisms or central control \citep{boons2017industrial}. An example of a practical question that could be answered is providing estimates for area development in case not all parameters are known. This could be useful for future development as well as when certain types of information are not public. The model has shown to provide reasonable parameter estimates for system level parameters.

The model will also be developed further methodologically by making it more realistic in several aspects. The next step is to make this model a data-driven model. Several variables, such as the input and output distributions, distance decay and transportation cost can be modeled or estimated on specific real-world data. Several examples of platforms are collecting information that could be used to inform the model. Examples vary from the life cycle analyses of one product~\citep{davis2009integration,wu2017agent}, to the functioning of an Eco-industrial park~\citep{jacobsen2006industrial}, to open data approaches to build an industrial symbiosis data repository~\citep{davis2017secondary,WebBen}. Such datasets, combined with the integration within a GIS (illustrated here by the EPRTR database integration) with interactive capabilities (e.g. an interactive web application%\comment JM{ (Shiny being a name of the brand of the R package, thus needs to be capitalized) "e.g. an application that could be coded in R language, for instance using the "Shiny" package" (or something of the likes, otherwise it's just a flashy app)} -> JR: I think we can remove the Shiny here as it is linked to implementation - and depending on the future developments of the project we may have computer scientists to build an application more ribust than a shiny one 
), should allow to both parametrize the model (initial configuration) and calibrate it (accuracy of the generated network of relations).

Furthermore, a refinement of economic processes included in the model is also an important research direction. The mechanism used to establish exchanges between companies is very simple and does not include behavioral considerations. A game theory framework with possibly more than two players is a possible way to take agents' behavior into account better as it has been done for cooperative relationships in a eco-industrial chain~\citep{li2012evolutionary}. The use of random utility models, suited in particular to model discrete choices~\citep{bierlaire1998discrete}, is another alternative. The economic structure is also relatively simplified. A potentially important model improvement could be a more realistic settings for company sizes, following e.g. a power-law for the distribution of sizes~\citep{simon1958size}. Furthermore, the model currently simulates processes on a short time scale, companies characteristics like position, size, type of activity, are assumed to be fixed. The integration into an evolutionary model on longer time scales~\citep{nelson2009evolutionary} would be relevant for longer term sustainable policies. Being more realistic on the location of companies based e.g. on real-world land use data and not only through urban density is also an important development, that we already suggested with the real-world illustration.

A key objective of future work will be to set the foundations for an open source application that can be used to monitor the circular economy, as well as create a marketplace for waste products. Data crowd sourcing, in interaction with feedback from models inspired from the one we developed, should make the actors aware of potentialities and foster exchanges between them.

\section*{Author contributions}
This study started as a project at the complex systems summer school of the Santa Fe Institute in 2016. All authors participated equally during the 4 week summer school. All authors participated in idea development, the model development, participated in writing and developing the study in general. JR conducted the programming and analyses. JB proposed the project. ES contributed significantly to the writing. Model development after the summer school mostly by JB, JR and MS. JB, JR, MS and JMS contributed substantially to the article's revisions.

\section*{Competing interests}

The authors declare no competing interests.

\section*{Data availability}

The datasets generated by simulation and analyzed during the current study are available from the dataverse repository, at \url{https://doi.org/10.7910/DVN/7XCWTN}.

Model code and results are available on the git repository of the project at\\
\url{https://github.com/SFICSSS16-CircularEconomy/CircularEconomy}.

\section*{Acknowledgments}

Results obtained in this paper were computed on the vo.complex-system.eu virtual organization of the European Grid Infrastructure ( http://www.egi.eu ). The authors thank the European Grid Infrastructure and its supporting National Grid Initiatives (France-Grilles in particular) for providing the technical support and infrastructure. This study started as a project at the complex systems summer school of the Santa Fe Institute.

%\bibliographystyle{elsarticle-harv.bst}
%\bibliographystyle{elsarticle-num.bst}
%\bibliography{biblio.bib}


%\appendix

\section*{Appendix A: Description of experiments}
\label{app:experiments}

\subsection*{Baseline grid search}

The baseline grid experiment to study model behavior was done with the following parameter values, with 100 repetitions each

\begin{itemize}
	\item $d_0 \in \{10, \ldots , 100\}$ with a step of 10
	\item $T_0 \in \{ 0.0, \ldots, 0.2\}$ with a step of 0.025
    \item $c \in \{0.5, \ldots, 3.5\}$ with a step of 0.25
    \item $\sigma \in \{0.01, \ldots, 0.1\}$ with a step of 0.01
    \item Setup type as uniform and synthetic city system
\end{itemize}

This gives a total of 23400 parameter points, corresponding to a total number of simulations of 2,340,000. This experiment corresponds to the result file \texttt{20180713{\_}210239{\_}DIRECTSAMPLING{\_}SYNTHETIC.csv} on the dataverse repository.

\subsection*{Influence of correlation level}

The targeted experiment to study the influence of the level of correlation was done on a synthetic city system, with the following parameter values and 100 repetitions each:

\begin{itemize}
	\item $d_0 \in \{10, \ldots , 90\}$ with a step of 20
	\item $T_0 \in \{0.0 ; 0.05 ; 0.1 ; 0.2\}$
    \item $c \in \{0.5, \ldots, 3.5\}$ with a step of 1.5
    \item $\sigma \in \{0.01, \ldots, 0.05\}$ with a step of 0.01
    \item $\alpha \in \{ 0.0 , \ldots , 15.0 \}$ with a step of 0.25
\end{itemize}

This gives a total of 18300 parameter points, corresponding to a total number of simulations of 1,830,000. This experiment corresponds to the result file \texttt{2018{\_}06{\_}19{\_}18{\_}50{\_}44{\_}DIRECTSAMPLING{\_}SYNTHETIC.csv} on the dataverse repository.

\subsection*{Genetic algorithm calibration}\label{app:ga}

The calibration with a Genetic Algorithm to find compromise solution aiming at minimizing total waste and relative cost was done with the standard NSGA2 algorithm implemented in OpenMole. A population of $\mu = 200$ individuals was taken, with a genome composed of parameters $(d_0,T_0)$ with boundaries $d_0 \in [1.0 ; 100.0]$ and $T_0 \in [0.0 ; 0.5]$. Objectives function were total waste and relative cost, and exogenous parameters (transportation cost and distribution width) were fixed at $c = 1.0$ and $\sigma = 0.1$. Setup was done on a synthetic city system for which meta-parameters were fixed at their default value. The algorithm was run following an island scheme with 1000 parallel islands. Negligible variations in the Pareto front were obtained after 50000 generations and the algorithm was stopped.

Several population files, including the one used corresponding to generation 50000, are available on the git repository at \url{https://github.com/SFICSSS16-CircularEconomy/CircularEconomy/tree/master/Models/Netlogo/netlogo6/explo/20180722_1631_NSGA2_SYNTHETIC_TRCOST3_DISTRIBSD0.01}.

\section*{Appendix B: Model behavior}

\subsection*{Statistical distributions}\label{app:distribs}

\begin{figure*}
	\includegraphics[width=\textwidth]{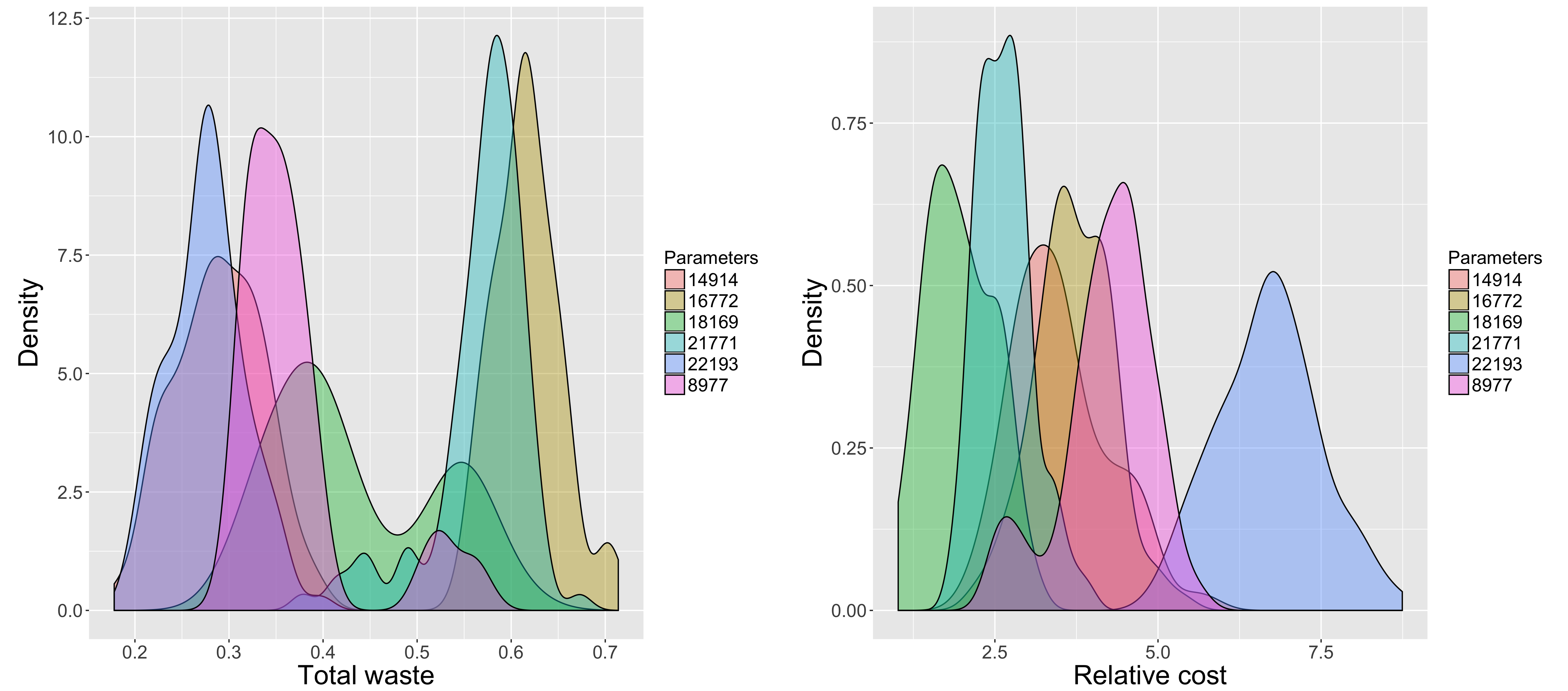}
	\caption{\textbf{Statistical distribution of indicators for some points in the parameter space.} We show for 6 different parameter values (given by parameter id in color, and which values are given in table~\ref{tab:distrib_params}) the statistical distribution of some indicators estimated on 100 repetitions of the model, for total waste and relative cost. We generally obtain clearly distinguishable distribution and significant tests for separating their averages as detailed in main text.}
\label{fig:stat-distrib}
\end{figure*}

Fig.~\ref{fig:stat-distrib} shows statistical distribution of some indicators for six sample points in the parameter space, that were chosen randomly among points of the baseline experiment. Their values are given below in Table~\ref{tab:distrib_params}.

%%%%%%%%%%%%%%%%%%
%%%%%%%%%%%%%%%%%%

%%%%%%%%%%%%
\begin{table}
	\centering
	\caption{\textbf{Parameter values for sample points}.}
    \label{tab:distrib_params}
	\begin{tabular}{|l|lllll|}
	\hline
	Id & $d_0$ & $\sigma$ & $T_0$ & $c$ & setup \\ \hline
    14914 & 100 & 0.07 & 0 & 1.5 & synthetic \\
    16772 & 90 & 0.02 & 0 & 2 & uniform \\
    18169 & 70 & 0.05 & 0.175 & 3 & synthetic \\
    21771 & 90 & 0.06 & 0 & 3 & uniform \\
    22193 & 60 & 0.1 & 0.1 & 1 & uniform\\
    8977 & 60 & 0.1 & 0.175 & 2.25 & uniform\\\hline
	\end{tabular}
\end{table}
%%%%%%%%%%%%

\subsection*{Indicators sensitivity}\label{app:sensitivity}

\begin{figure*}
	\includegraphics[width=\textwidth]{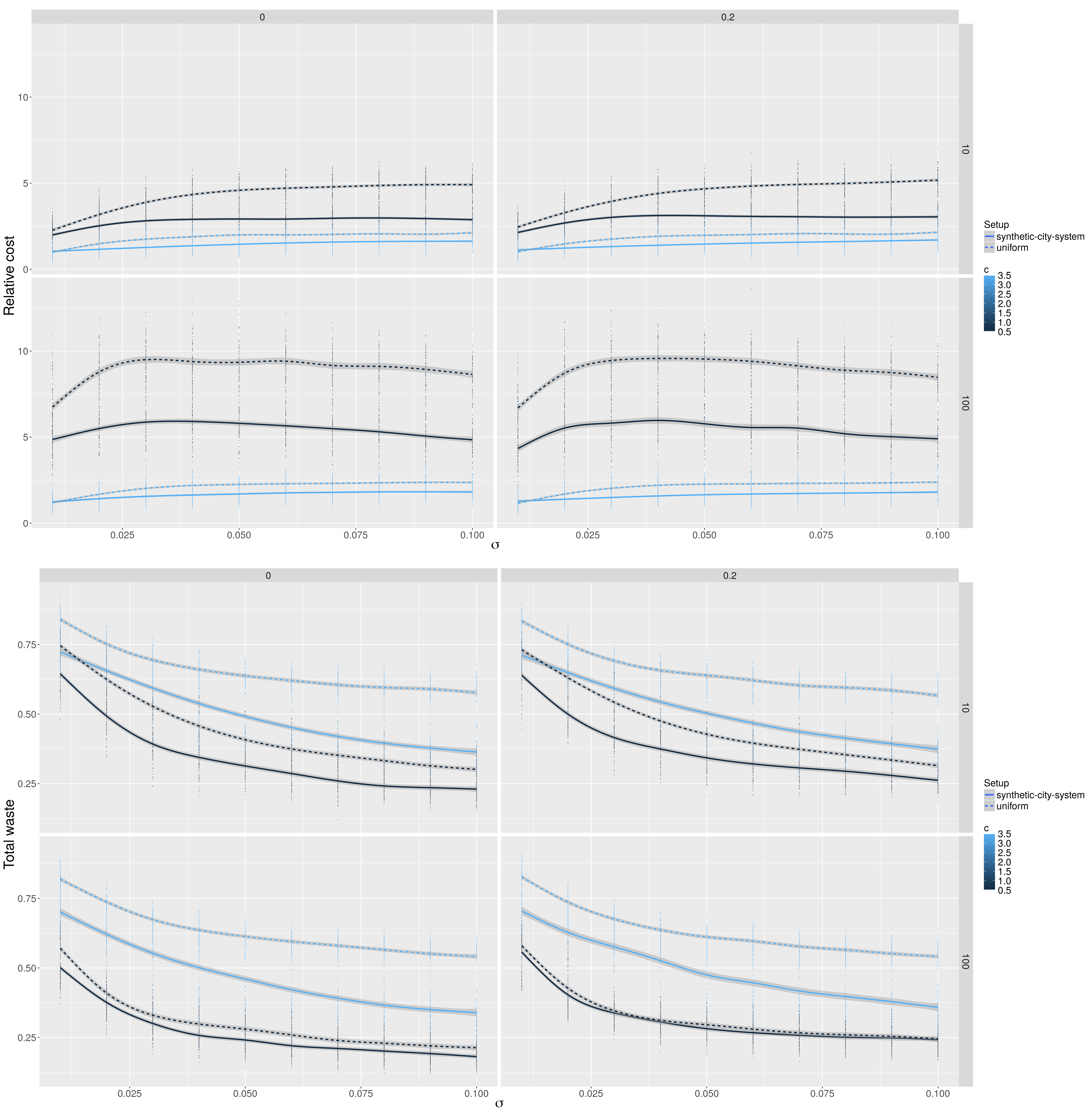}
	\caption{\textbf{Sensitivity of indicators.} We show the variation of total waste and relative cost, for a sample from the exploration of the parameter space for the baseline model (first experiment), for uniform and synthetic city system setup type, and varying values of $d_0$ (rows) and of $T_0$ (columns).}
\label{fig:indics}
\end{figure*}

We give the behavior of waste and cost indicators in Fig.~\ref{fig:indics}. Dotted points correspond to single runs and smoothed lines to statistical averages. We show the curves for extreme parameter values only to ensure readability. This gives a broad idea of the diversity of regimes the model can produce when varying the parameters: low cost and low waste, high cost and high waste, but also high cost and low waste for example.

We clearly distinguish the discrepancy between a uniform and a synthetic city system. We obtain some expected qualitative behaviors such as (i) a decrease of waste when $\sigma$ increase; (ii) an increase of waste when transportation cost increases; (iii) an increase of relative cost when $c$ decrease. We also obtain unexpected behaviors, such as (i) an increase of waste with $T_0$ for low transportation costs; (ii) a slight u-shape behavior of cost as a function of $\sigma$ for high $d_0$ and low $c$; (iii) an effect of $d_0$ on waste which leads to no difference between setup types for high values of $d_0$.

Note that the rare parameter settings giving bimodal values correspond to high values of $\sigma$ and $c$, are not specifically considered here (for example by fitting separately the two modes) as we plot indicators as a function of $\sigma$. This aspect is however important as a function of $\alpha$ as detailed in main text.

%%%%%%%%%%%%%%%%%%
%%%%%%%%%%%%%%%%%%

%%%%%%%%%%%%%
\section*{Appendix C: Variation of Pareto fronts for policy optimization}\label{app:fronts}

\begin{figure*}
	\includegraphics[width=\textwidth]{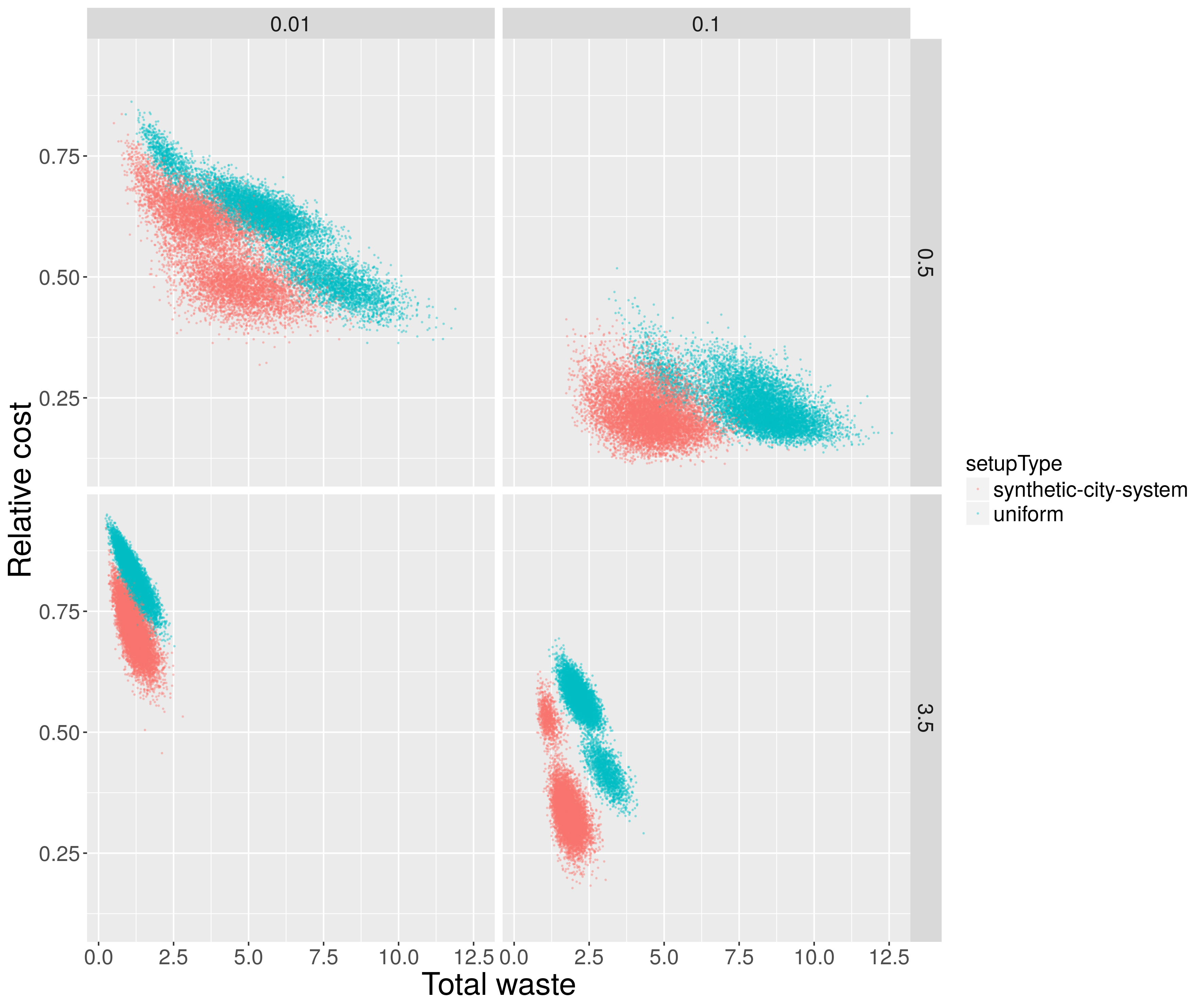}
	\caption{\textbf{Points cloud of total waste against relative cost, exhibiting Pareto fronts.} Several fronts are shown, each at extreme fixed values of transportation cost $c$ (rows) and distribution standard deviation $\sigma$ (columns), color gives the type of setup (uniform or synthetic city system).}
\label{fig:pareto}
\end{figure*}

We show in Fig.~\ref{fig:pareto} the point clouds for cost and waste, at extreme fixed values of $\sigma$ and $c$, for the different setup types. These exhibit Pareto fronts on their boundaries. We obtain that synthetic city systems are always fully dominating uniform configurations, what was expected as companies are clustered in space within the city system setup. We visualize here the variations in the position of Pareto fronts: at fixed low $c$, higher $\sigma$ values improve cost but not waste; at fixed high $c$, higher $\sigma$ values improve cost but degrade waste performance; at fixed low $\sigma$, lower $c$ values improve cost but degrade waste also; and at fixed high $\sigma$, higher $c$ values improve both indicators.

%%%%%%%%%%%%%%%%%%
%%%%%%%%%%%%%%%%%%

\section*{Appendix D: Illustration of real world application at a local scale}

\begin{figure*}
\centering
\includegraphics[width=0.65\textwidth]{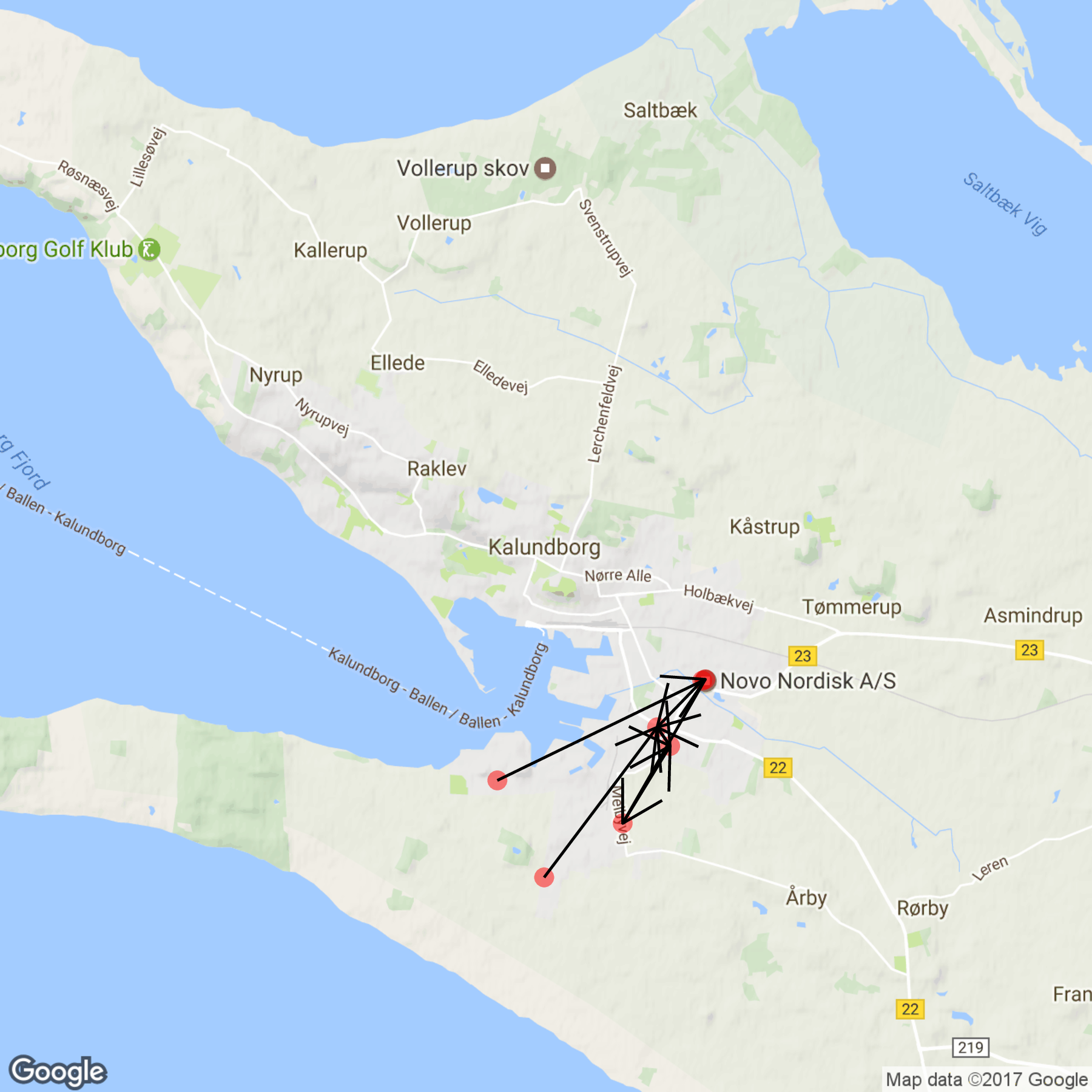}
\caption{Example real world deterministic location, using the geolocation of 6 companies located in Eco-industrial park Kalundborg, Denmark}
\label{fig:RealExam}
\end{figure*}

We also illustrate a potential application of the model at a very large scale. In Fig.~\ref{fig:RealExam} we have implemented the model in a real world setting corresponding to the geographical span of an eco-industrial park. The simplified model can be run on this real world deterministic setting, taking a list of company names and the spatial correlation as the input. The real world deterministic setting uses the same variables as the model described above. The difference with the model described above is that the model takes a list of company names as an input. The google API locates the companies on the map and uses the actual geographical distances to run the model. So the parameter $d_{ij}$ is now a fixed parameter describing the distance between two companies according to the google API. For illustration purposes the we use 6 existing companies in the Eco-industrial park of Kalundborg, Denmark. The deterministic application can be used to describe or compare different areas on their geographical characteristics. In the example the input and output distributions are assigned randomly, but as mentioned in main text, future versions of the model can be more data-driven as the EPRTR application, and improve the model beyond theoretical purposes. By informing the parameters with data the model can potentially be used to serve as a practical tool. This way, the deterministic implementation serves as an outlook for future developments of ABMs in the field of industrial ecology. We did not systematically explore results for this model, however, the code and shiny implementation can be found on the Github page dedicated to this paper described above. 

%%%%%%%%%%%%%%
%%%%%%%%%%%%%%%


%\begin{thebibliography}{00}

%\end{thebibliography}

\end{document}